\documentclass{article}

\usepackage{PRIMEarxiv}

\usepackage[utf8]{inputenc} % allow utf-8 input
\usepackage[T1]{fontenc}    % use 8-bit T1 fonts
\usepackage{hyperref}       % hyperlinks
\usepackage{url}            % simple URL typesetting
\usepackage{booktabs}       % professional-quality tables
\usepackage{amsfonts}       % blackboard math symbols
\usepackage{nicefrac}       % compact symbols for 1/2, etc.
\usepackage{microtype}      % microtypography
\usepackage{lipsum}
\usepackage{fancyhdr}       % header
\usepackage{apacite}
\usepackage{tabularx}
\usepackage{graphicx}       % graphics
\graphicspath{{media/}}     % organize your images and other figures under media/ folder

\title{How Jungian Cognitive Functions Explain MBTI Type Prevalence in Computer Industry Careers
}

\usepackage{authblk} % For author affiliations

\author{
\textbf{Arya VarastehNezhad}\textsuperscript{1} \quad
\textbf{Behnam Agahi}\textsuperscript{2} \quad
\textbf{Soroush Elyasi}\textsuperscript{3} \quad
\textbf{Reza Tavasoli}\textsuperscript{4} \quad
\textbf{Hamed Farbeh}\textsuperscript{2}\thanks{Corresponding author: Hamed Farbeh, farbeh@aut.ac.ir.}  \\ \vspace{0.5em} % Add some space before affiliations
\textsuperscript{1}Department of Computer Engineering, University of Tehran, Tehran, Iran \\
\textsuperscript{2}Department of Computer Engineering, Amirkabir University of Technology, Tehran, Iran \\
\textsuperscript{3}School of Computing and Engineering, University of West London, London, UK \\
\textsuperscript{4}Department of Computer Science and Engineering, University of South Carolina, Columbia, USA \\ \vspace{0.5em} % Add some space before email
\texttt{aryavaraste@ut.ac.ir, behnamagahi49@aut.ac.ir, soroush.elyasi@uwl.ac.uk, tavasoli@email.sc.edu, farbeh@aut.ac.ir}
}

\begin{document}
\maketitle

\begin{abstract}
This study investigates the relationship between Carl Jung's cognitive functions and success in computer industry careers by analyzing the distribution of Myers-Briggs Type Indicator (MBTI) types among professionals in the field. Building on Carl Jung's theory of psychological types, which categorizes human cognition into four primary functions, Sensing, Intuition, Thinking, and Feeling, this study investigates how these functions, when combined with the attitudes of Extraversion and Introversion, influence personality types and career choices in the tech sector. Through a comprehensive analysis of data from 30 studies spanning multiple countries and decades, encompassing 18,264 individuals in computer-related professions, we identified the most prevalent cognitive functions and their combinations. After normalizing the data against general population distributions, our findings showed that individual Jungian functions (Te, Ni, Ti, Ne), dual function combinations (Ni-Te, Ti-Ne, Si-Te, Ni-Fe), and MBTI types (INTJ, ENTJ, INTP, ENTP, ISTJ, INFJ, ESTJ, ESTP) had significantly higher representation compared to general population norms. The paper addresses gaps in the existing literature by providing a more nuanced understanding of how cognitive functions impact job performance and team dynamics, offering insights for career guidance, team composition, and professional development in the computer industry, and a deeper understanding of how cognitive preferences influence career success in technology-related fields.
\end{abstract}

% keywords can be removed

\section{Introduction}

Carl Gustav Jung, a Swiss psychiatrist and psychoanalyst, developed a theory of psychological types that provides profound insights into human personality, the key characteristics and emphasis of which are shown in Table \ref{tab:table1}. Central to Jung's theory are the concepts of Extraversion and Introversion, which describe the general attitude or orientation of an individual's energy. These attitudes form the basis for how individuals interact with the world and other people. Building on these attitudes, Jung introduced four primary cognitive functions, Sensing, Intuition, Thinking, and Feeling, to describe how individuals perceive information and make decisions. These functions, when combined with the attitudes of Extraversion and Introversion, form the core of Jung's personality theory, and Table \ref{tab:table1} provides an overview of key characteristics and emphasis of different functions \cite{jung2016psychological, jung2014structure, jung1971psychological}.   

Extraversion and Introversion are two opposing attitudes that describe the direction in which an individual’s energy flows. Perceiving functions are concerned with how we gather and interpret information from the environment. Jung identified two key perceiving functions: Sensing (S) and Intuition (N). Judging functions are concerned with how we evaluate information and make decisions. Jung identified two key judging functions: Thinking (T) and Feeling (F) \cite{jung1971personality}.

\begin{table} [htbp]
 \caption{Overview of Jungian Cognitive Functions and Attitudes \protect\cite{jung2016psychological, jung1971personality, drenth2014my}}   
  \centering
  {\fontsize{8}{9}\selectfont
  \begin{tabular}{
  p{1.2cm}
  p{1.5cm}
  p{2cm}
  p{1.6cm}
  p{5.4cm}
  p{2.3cm}
  }

    \toprule
    \cmidrule(r){1-2}
    Category     & Function     & Focus
    & Manifestation &  Key Characteristics & Emphasis \\
    \midrule
    Attitudes & Extraversion & Directing energy outward& N/A & Sociable, action-oriented, comfortable in groups, energized by external stimuli.& Engagement with the external world\\
     & Introversion& Directing energy inward& N/A& Reserved, prefer solitude and reflection, energized by internal thoughts and ideas.& Introspection and internal processing\\

     Perceiving Functions & Sensing (S) & Concrete experiences through the five senses & Extraverted Sensing (Se) & Detail-oriented, spontaneous, highly responsive to immediate environment. & Immediate sensory input \\

     &  &  & Introverted Sensing (Si) & Strong sense of tradition, ability to recall detailed information, connects current experiences with past memories. & Comparing current experiences to past ones\\

     & Intuition (N) & Abstract patterns, possibilities, and the future & Extraverted Intuition (Ne) & Inventive, curious, excellent at brainstorming, explores multiple possibilities and connections. & Generating diverse ideas\\

      &  &  & Introverted Intuition (Ni) & Insightful, future-oriented, excels at understanding complex patterns and long-term outcomes, synthesizes information into a cohesive vision. & Developing a cohesive, internal vision \\

      Judging Functions & Thinking (T) & Logic, objectivity, and analysis & Extraverted Thinking (Te) & Decisive, task-oriented, skilled at implementation and resource management, organizes the external world efficiently. & External organization and efficiency\\

       &  &  & Introverted Thinking (Ti) & Introspective, precise, excels at developing complex theories, analyzes information to create an internal framework of understanding. & Internal coherence and understanding\\

        & Feeling (F) & Values, empathy, and human relationships & Extraverted Feeling (Fe) & Empathetic, sociable, skilled at building relationships, fosters cooperation, aligns with others' values to maintain harmony. & Communal values, social harmony, and cooperation\\

        &  &  & Introverted Feeling (Fi) & Introspective, authentic, guided by a strong moral compass, aligns actions with one's deeply held values. & Personal values, authenticity, and integrity\\
    \bottomrule
  \end{tabular}
  }
  \label{tab:table1} % Unique label
\end{table}

\begin{figure} [htbp]
  \centering
    \includegraphics[width=0.3\textwidth]{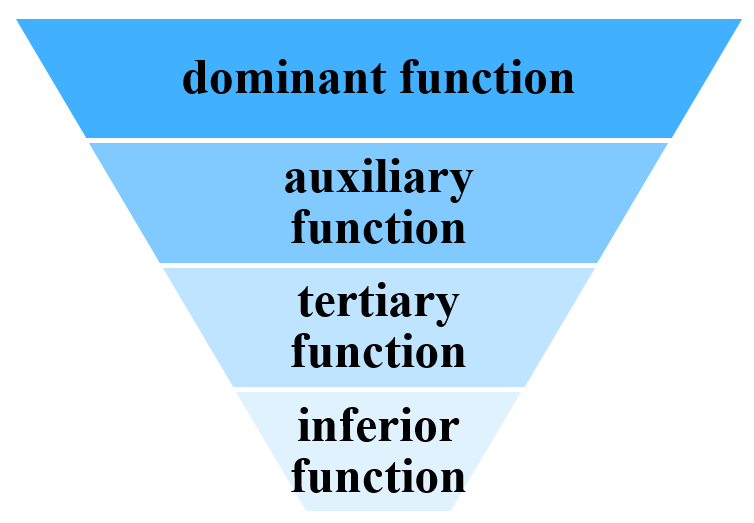} % adjust path and size
  \caption{Hierarchy of Jungian Cognitive Functions}
  \label{fig:fig1} % Unique label
\end{figure}

Jung proposed that each individual has a unique combination of cognitive functions, organized into a hierarchy that shapes their personality (Figure \ref{fig:fig1}). This hierarchy consists of the dominant function, which is the most developed and influential, shaping the individual’s primary mode of interaction with the world. The auxiliary function serves as the secondary function, supporting and balancing the dominant function by providing additional perspective and flexibility. The tertiary function, though less developed, begins to emerge more prominently as the individual matures, contributing to their overall personality. Finally, the inferior function is the least developed, often residing in the unconscious, and can manifest in less mature forms, presenting opportunities for personal growth \cite{jung2016psychological, drenth201316}.

The Myers-Briggs Type Indicator (MBTI) \cite{myers1985manual} is a widely used personality assessment tool that categorizes individuals into one of sixteen distinct personality types. Developed by Katharine Cook Briggs and her daughter Isabel Briggs Myers in the mid-20th century, the MBTI is based on Carl Jung’s theory of psychological types, particularly his concept of cognitive functions. The MBTI makes Jung’s complex theories more accessible and practical, allowing individuals to better understand themselves and others in various contexts, including personal growth, relationships, and career development by using four dichotomies to categorize personality preferences, resulting in a four-letter code. These dichotomies are Extraversion (E) vs. Introversion (I), Sensing (S) vs. Intuition (N), Thinking (T) vs. Feeling (F), and Judging (J) vs. Perceiving (P). Each represents a fundamental preference in how individuals direct energy, gather information, make decisions, and approach the external world \cite{drenth2014my}.    

The E/I dichotomy describes energy direction. Extraverts are energized by the external world of people and activities, being outgoing and action-oriented. Introverts, conversely, are energized by their internal world of thoughts and reflection, tending to be reserved and introspective. The S/N dichotomy reflects information gathering. Sensing individuals rely on their five senses, focusing on concrete details and practical realities. Intuitive individuals focus on patterns, possibilities, and abstract connections, drawn to big-picture thinking and innovation. The T/F dichotomy describes decision-making. Thinking individuals prioritize logic, objectivity, and consistency, striving for fairness and truth. Feeling individuals prioritize values, empathy, and the impact on people, seeking harmony and aligning with values. The J/P dichotomy describes lifestyle preferences. Judging individuals prefer structure, order, and planning. Perceiving individuals prefer flexibility, adaptability, and keeping options open \cite{pittenger1993utility}.    

MBTI is deeply rooted in Carl Jung's theory of psychological types, which categorizes human cognition into four primary functions. Each of the 16 MBTI personality types is defined by a unique combination of four of these Jungian functions. These functions are ordered in a hierarchy, where the dominant function is the most influential in a person's cognitive processes, followed by the auxiliary, tertiary, and inferior functions. The interaction of these functions within each MBTI type shapes the individual's approach to perceiving the world and making decisions, which in turn influences their compatibility with various career paths, including those in computer-related fields. Table \ref{tab:table2} outlines the four Jungian functions for each of the 16 MBTI types.

\begin{table} [htbp]
 \caption{Jungian Function Stacks for Each MBTI Type }
  \centering
  {\fontsize{8}{10}\selectfont
  \begin{tabular}{
  p{2.1cm}
  p{0.4cm}
  p{0.4cm}
  p{0.4cm}
  p{0.4cm}
  p{0.4cm}
  p{0.4cm}
  p{0.4cm}
  p{0.4cm}
  p{0.4cm}
  p{0.4cm}
  p{0.4cm}
  p{0.4cm}
  p{0.4cm}
  p{0.4cm}
  p{0.4cm}
  p{0.4cm}
  }

    \toprule
    \cmidrule(r){1-2}

    & ISTJ & ISFJ & INTJ & INFJ & ISTP & INTP & ISFP & INFP & ESTP & ESFP & ENTP & ENFP & ESTJ & ENTJ & ESFJ & ENFJ\\
    \midrule
    Dominant function  & Si	& Si & Ni & Ni & Ti & Ti & Fi & Fi & Se & Se & Ne & Ne & Te & Te & Fe & Fe\\
    Auxiliary function & Te & Fe & Te & Fe & Se & Ne & Se & Ne & Ti & Fi & Ti & Fi & Si & Ni & Si & Ni\\
    Tertiary function  & Fi	& Ti & Fi &	Ti & Ni & Si & Ni & Si & Fe & Te & Fe & Te & Ne & Se & Ne & Se\\
    Inferior function  & Ne & Ne & Se & Se & Fe & Fe & Te & Te & Ni & Ni & Si & Si & Fi & Fi & Ti & Ti\\

    \bottomrule
  \end{tabular}
  }
  \label{tab:table2} % Unique label
\end{table}

The computer industry, characterized by rapid technological advancements and diverse roles, presents a unique landscape where understanding personality types and cognitive functions can be valuable. The field encompasses a wide range of positions, from highly technical roles like software engineering and data science to more interpersonal roles such as project management and user experience design. Each of these roles may benefit from different cognitive strengths and work styles, making the application of MBTI and Jung's cognitive functions insightful for career guidance and team composition.

In the context of the computer industry, MBTI types may correlate with certain career preferences and strengths. For instance, INTJ (Introverted, Intuitive, Thinking, Judging) types might excel in roles requiring strategic thinking and complex problem-solving, such as systems architecture. Conversely, ENFJ (Extraverted, Intuitive, Feeling, Judging) types might thrive in roles that involve team leadership, communication, and people management, like project management or technical team leadership. Understanding these potential alignments can help individuals make informed career choices and assist organizations in building well-balanced teams \cite{berens2001quick, sach2010use}.    

Jung's cognitive functions offer a more nuanced perspective on how individuals process information and make decisions, which is crucial in the fast-paced and often complex environment of the computer industry. For example, those with strong Introverted Thinking (Ti) might excel in roles requiring deep analytical skills and logical problem-solving, such as algorithm development or data analysis. Those with developed Extraverted Feeling (Fe), which is the auxiliary function for ENFJs, might be well-suited for roles involving team coordination, stakeholder management, or user advocacy in areas like agile project management or customer success in tech companies.

The relevance of MBTI and Jung's functions extends beyond individual career choices, significantly influencing team dynamics, organizational structure, and ultimately, career trajectories within the dynamic computer industry. Understanding the cognitive diversity within a team, such as balancing detail-oriented Sensing types with big-picture Intuitive types, or leveraging both Thinking types' logical rigor and Feeling types' focus on user needs and harmony, can foster more effective collaboration, communication, and problem-solving. This understanding informs crucial decisions in hiring, team formation, and management strategies tailored to tech organizations \cite{barrick1991big}. Indeed, the frequent career path changes and misplacements common in the industry highlight a critical need for a better grasp of personality-job alignment. While studies indicate that companies leveraging personality assessments see benefits like higher employee satisfaction and lower turnover \cite{giannoukou2023role, hackstonpersonality}, a deeper, more systematic investigation into how specific Jungian cognitive functions manifest and impact the unique demands of computer industry roles is warranted. Therefore, this paper aims to fill this gap by exploring the specific role of Carl Jung's cognitive functions and derived MBTI types within computer industry careers. We seek to provide a more comprehensive and nuanced understanding by examining not only the prevalence of different personality profiles in tech roles but also by investigating the critical alignment between specific cognitive functions and job requirements. To guide this focused investigation, we address the following key research questions:

\begin{enumerate}
    \item What is the distribution of MBTI personality types and Jungian cognitive functions, and combinations of Jungian cognitive functions (e.g., Ni-Te, Ti-Ne) in computer industry careers, and how does this distribution differ from the general population?

    \item Beyond prevalence, what are the potential implications of specific Jungian cognitive function profiles for job performance, team dynamics, and career development within the computer industry?
\end{enumerate}

\subsection{Structure of the Paper}
The remainder of this paper is organized as follows: Section \ref{sec:headings} provides an overview of the relevant literature on personality types, cognitive functions, and their applications within the technology sector, establishing the context for our investigation and highlighting existing research gaps. Section \ref{sec:methodology} details the methodology employed in this study, including data collection from multiple sources, aggregation of MBTI type distributions, and the normalization process used to compare these distributions against general population norms. In Section \ref{sec:prevalence}, we present the results of our analysis, focusing on the prevalence of individual Jungian functions, function combinations, and MBTI types within the computer industry. Section \ref{sec:discussion} discusses the implications of our findings for recruitment, team composition, and organizational culture and analyzes the results by elaborating tables and detailed explanations. Finally, Section \ref{sec:challenges} offers concluding remarks and outlines directions for future research, acknowledging limitations of the current study and suggesting avenues for further exploration.

\section{Related Work}
\label{sec:headings}

The intersection of personality types and job performance in the tech sector has been explored through various lenses, particularly focusing on the MBTI and other personality assessment tools. Here, we review several recent studies that highlight the role of personality of people in the computer industry, identifying key findings and existing research gaps.

\citeA{mcpherson2007students} explored the relationship between personality types and the choice of major among information technology students. They found significant correlations between MBTI types and students' preferences for specific tech majors, suggesting that personality assessments can guide educational and career choices in the tech field. \citeA{lounsbury2007investigation} investigated the relationship between personality traits and job satisfaction among IT professionals. The study identified traits such as assertiveness, emotional resilience, and teamwork disposition as significant predictors of job and career satisfaction, indicating the relevance of personality traits in understanding job performance in the tech sector. \citeA{mirza2015generating} focused on generating personalized job role recommendations for the IT sector by analyzing personality traits using Holland Codes and the Five Factor Model. Their study highlighted the direct correlation between personality traits and job role success, emphasizing the importance of personality assessments in career guidance. \citeA{elyasi2023mbti} explored personality types among software engineers in Iran. Their study finds that personality types such as ISTJ, INTJ, ESTJ, and ENTJ are more prevalent, while ISFJ, ISFP, ESFP, ENFP, and ESFJ are less common.

\citeA{yang2022research} discusses the application of MBTI in corporate settings and their findings suggest that MBTI is instrumental in recruitment and team-building processes, aiding in better talent screening and enhancing team harmony and efficiency, which are essential for tech companies aiming to optimize team performance. \citeA{montequin2012using} provide evidence that understanding personality traits can improve team outcomes in IT projects. Their study highlights that MBTI can help form balanced teams by considering the complementarity of different personality types, fostering better collaboration and project success. \citeA{iqbal2019predicting} propose a systematic mapping of MBTI traits to software development tasks. This approach ensures that team members' cognitive strengths are aligned with their job roles, enhancing the overall quality and efficiency of software development processes. Lastly, incorporating cognitive functions provides a more stable and consistent framework for personality assessment. \citeA{sendall2015longitudinal} suggest that understanding the deeper cognitive processes behind personality traits can lead to more reliable assessments and better long-term predictions of job performance.

\subsection{Gaps in the Literature}

Despite extensive research examining the relationship between personality types and job performance in the tech sector, several critical gaps persist, limiting both our theoretical understanding and practical application of these findings. A primary limitation lies in the predominantly correlational nature of existing studies linking MBTI types (and other personality indicators) to career choices and job satisfaction. While these studies consistently identify patterns, such as the prevalence of ISTJ types in software engineering roles, they fail to determine the causal mechanisms underlying these relationships. This superficial understanding limits our ability to leverage personality insights effectively in recruitment, team composition, and professional development.

Current research also suffers from methodological shortcomings in its treatment of personality frameworks. Most studies rely on broad MBTI categories, overlooking the nuanced cognitive functions that form their theoretical foundation. While Jung's cognitive functions (Te, Ni, Ti, Ne) offer a more granular framework for understanding individual differences, research rarely examines how these specific functions, independently or in combination, influence performance in technical roles. This gap is particularly problematic when addressing team dynamics and individual performance issues, where understanding specific cognitive preferences could inform targeted interventions.

Another significant limitation is the lack of normalized data in personality type distribution studies. Researchers often present raw frequencies or percentages of MBTI types without contextualizing these findings against baseline population distributions. This oversight makes it hard to determine whether certain personality types are genuinely overrepresented in technical roles or merely reflect broader demographic patterns. The absence of standardized normalization procedures also prevents meaningful cross-study comparisons and meta-analyses. Finally, existing research lacks the comprehensive scope needed to fully understand personality dynamics across the diverse and rapidly evolving tech sector. Many studies draw from limited samples, often restricted to specific companies, geographic regions, or educational institutions, potentially introducing significant sampling bias. Moreover, as the technology industry continues to create new roles and specializations, there is an urgent need for research that captures these emerging contexts and their unique personality-performance dynamics.

\section{Methodology}
\label{sec:methodology}

\subsection{Data Gathering}

This section details the data-gathering process used to analyze the distribution of MBTI types in computer-related fields, utilizing the articles from the literature. Each article provided percentages of various MBTI types within their studied populations, along with the total population sizes. This comprehensive dataset formed the basis of our analysis. The articles from the literature are shown in Table \ref{tab:table3}. The diversity of sources ensures a robust and comprehensive understanding of MBTI distribution across different regions and time periods. However, it is important to note that there is still work to be done to gather data from countries that have not yet been studied, which could provide further insights and validate our findings across a broader spectrum.

\begin{table} [htbp]
 \caption{Raw MBTI Type Distribution Data from Literature Review}
  \centering
  \vspace{0.2cm}
  \makebox[\textwidth][c]{%
  {\fontsize{7}{10}\selectfont
  \begin{tabular}{
  p{4cm}
  p{0.35cm}
  p{0.35cm}
  p{0.9cm}
  p{0.24cm}
  p{0.24cm}
  p{0.24cm}
  p{0.24cm}
  p{0.24cm}
  p{0.24cm}
  p{0.24cm}
  p{0.24cm}
  p{0.24cm}
  p{0.24cm}
  p{0.24cm}
  p{0.24cm}
  p{0.24cm}
  p{0.24cm}
  p{0.24cm}
  p{0.24cm}
  }

    \toprule
    \cmidrule(r){1-2}

    Source & Year &	Count &	Country & ISTJ & ISFJ & INTJ & INFJ & ISTP & INTP & ISFP & INFP & ESTP & ESFP & ENTP & ENFP & ESTJ & ENTJ & ESFJ & ENFJ\\
    \midrule

    \citeA{elyasi2023mbti} &
    2023&	102&	Iran&	19.61&	2.94&	3.92&	14.7&	7.84&	2.94&	2.94&	14.7&	3.92&	1.96&	0.98&	3.92&	11.76&	0.98&	0&	6.86\\

    \citeA{perez2018investigating} &
    2018&	1252&	Canada&	19.5&	3.3&	3&	10.1&	8.2&	2.9&	4.3&	9.9&	5.4&	2.4&	3.6&	6.8&	10.9&	2.5&	2.3&	5\\

    &
    & 100&	Brazil&	19.12&	2.94&	1.47&	7.35&	4.41&	4.41&	2.94&	13.24&	11.76&	1.47&	2.94&	7.35&	11.76&	2.94&	1.47&	4.41\\

    &
    & 53&	Mexico&	15.09&	3.77&	7.55&	1.89&	16.09&	9.43&	9.43&	9.43&	5.66&	0&	3.77&	3.77&	3.77&	3.77&	5.66&	1.89\\

    \citeA{gilal2017effective} &
    2017&	75&	MC&	66&	18.7&	0&	23&	17&	11&	0&	13.3&	18.7&	0&	0&	20.6&	5.6&	5.6&	0&	0\\

    \citeA{capretz2015influence} &
    2015&	100&	Cuba&	10&	7&	1&	6&	5&	2&	1&	5&	15&	6&	3&	2&	25&	2&	3&	7\\

    \citeA{omar2015assessing} &
    2015&	81&	Malaysia&	12.35&	6.17&	6.64&	19.75&	2.47&	2.47&	4.94&	7.41&	2.47&	1.23&	0&	1.23&	9.88&	4.94&	2.47&	13.58\\

    \citeA{dargis2015relationship} &
    2015&	35&	Latvia&	0&	22.86&	8.57&	0&	5.71&	8.57&	5.71&	0&	2.86&	0&	2.86&	0&	0&	14.29&	25.71&	2.86\\

        \citeA{raza2014personality} &
    2014&	110&	Pakistan&	14&	9&	3&	7&	6&	7&	4&	8&	4&	6&	5&	11&	9&	4&	2&	1\\

        \citeA{okike2014problem} &
    2014&	29&	Nigeria&	17.24&	6.9&	6.9&	0&	6.9&	3.45&	3.45&	10.34&	3.45&	3.45&	6.9&	0&	13.79&	3.45&	3.45&	10.34\\

        \citeA{varona2014comparison} &
    2014&	100&	Cuba&	7&	7&	6&	2&	4&	3&	5&	2&	13&	5&	7&	13&	8&	8&	4&	6\\

        \citeA{raza2015personality} &
    2012&	70&	Pakistan&	34&	9&	13&	7&	15&	9&	10&	23&	2&	4&	17&	13&	23&	6&	0&	15\\

        \citeA{varona2011personality} &
    2011&	103&	Cuba&	9.71&	6.8&	0.97&	5.8&	4.85&	1.94&	0.97&	5.8&	13.6&	5.8&	3.88&	1.94&	26.2&	1.94&	2.91&	6.8\\

        \citeA{ahmed2010learning} &
   2010&	235&	N/A&	28.2&	9.4&	2.35&	7.1&	3.53&	3.53&	4.71&	4.7&	3.53&	3.5&	2.35&	2.35&	2.35&	2.35&	2.35&	17.6\\

        \citeA{schaubhut2008mbti} &
    2008&	5528&	MC&	54.4&	13.4&	6.8&	23.3&	22.9&	7.9&	13.6&	27.7&	15.3&	6.4&	13&	22.2&	31.3&	11.3&	5.5&	19.8\\

        \citeA{capretz2015psychological} &
    2008&	68&	Brazil&	19&	2.9&	1.47&	7.4&	4.41&	4.41&	2.94&	13&	11.7&	1.5&	2.94&	7.35&	11.7&	2.94&	1.47&	4.41\\

        \citeA{choi2008exploring} &
    2008&	128&	USA&	21.1&	5.5&	0.8&	3.9&	3.1&	6.3&	7&	7&	5.5&	5.5&	5.5&	3.1&	14.8&	5.5&	2.3&	3.1\\

        \citeA{galpin2007learning} &
    2007&	72&	S. Africa&	17.7&	12.9&	1.2&	4.7&	1.2&	2.4&	2.4&	4.7&	4.7&	3.5&	4.7&	3.5&	18.8&	8.4&	5.9&	2.4\\

        \citeA{choianalysis} &
    2006&	128&	N/A&	21&	5.5&	0.8&	3.9&	3.1&	6.3&	7&	7&	5.5&	5.5&	5.5&	3.1&	14.8&	5.5&	2.3&	2.38\\

        \citeA{miller2004cognitive} &
    2004&	33&	N/A&	21.2&	15&	3.03&	6.1&	6.06&	3.03&	3.03&	3&	0&	0&	0&	6.06&	18.2&	9.09&	0&	6.06\\

        \citeA{karn2005study} &
    2004&	19&	N/A&	0&	5.3&	15.8&	0&	0&	0&	0&	16&	0&	0&	5.26&	5.26&	5.26&	0&	42.1&	5.26\\

        \citeA{kaluzniacky2004managing} &
    2004&	66&	N/A&	39&	4.5&	0&	12&	6&	0&	4.5&	1.5&	1.5&	0&	1.5&	4.5&	18&	3&	0&	3\\

        \citeA{capretz2003personality} &
    2003&	100&	N/A&	24&	3&	0&	7&	8&	5&	0&	8&	8&	1&	3&	7&	15&	4&	1&	4\\

        \citeA{capretz2002software} &
    2002&	1252&	Canada&	19.5&	3.3&	3&	10.1&	8.2&	2.9&	4.3&	9.9&	5.4&	2.4&	3.6&	6.8&	10.9&	2.5&	2.3&	5\\

        \citeA{varona2012evolution} &
    2002&	419&	N/A&	21.7&	5.5&	2.15&	8.1&	6.21&	0.95&	2.15&	6.4&	5.49&	0.95&	2.63&	8.35&	17.4&	3.58&	3.34&	5.01\\

        \citeA{teague1998personality} &
    1998&	38&	N/A&	5.26&	2.6&	0&	11&	7.89&	0&	7.89&	5.3&	5.2&	2.63&	7.89&	15.8&	10.5&	2.63&	13.2&	2.63\\

            \citeA{bradley1997effect} &
    1997&	22&	N/A&	18.2&	4.5&	4.55&	23&	4.55&	4.5&	4.55&	0&	0&	0&	0&	0&	0&	9.09&	9.09&	18.2\\

            \citeA{turley1995competencies} &
    1995&	59&	USA&	10.16&	5.08&	6.77&	6.77&	8.47&	6.77&	6.77&	10.16&	0&	3.39&	3.39&	6.77&	8.47&	3.39&	6.77&	5.77\\

            \citeA{carland1990cognitive} &
    1990&	92&	USA&	1.09&	1.09&	1.09&	1.09&	7.61&	15.22&	15.22&	3.26&	4.35&	19.57&	11.96&	6.52&	1.09&	3.26&	6.52&	1.09\\

            \citeA{thomsett1980people} &
    1990&	656&	N/A&	38&	5.2&	2.9&	6.5&	0.6&	0.6&	0.5&	0.6&	0.6&	1.5&	1.5&	1.7&	25&	4.9&	3.8&	6\\

            \citeA{smith1989personality} &
    1989&	37&	N/A&	35.1&	8.1&	0&	8.1&	2.7&	0&	0&	5.4&	2.7&	0&	0&	2.7&	29.7&	2.7&	0&	0\\

            \citeA{buie1988psychological} &
   1988&	47&	USA&	19.2&	4.3&	8.5&	12.8&	8.5&	0&	6.4&	14.9&	2.1&	6.4&	4.3&	0&	6.4&	0&	2.1&	4.3\\

            \citeA{westbrook1988frequencies} &
    1988&	153&	N/A&	19.6&	3.3&	0&	6.5&	5.9&	3.9&	0.7&	4.6&	3.9&	0.7&	5.2&	5.9&	27.5&	2.6&	1.3&	8.5\\

            \citeA{lyons1985dp} &
    1985&	1229&	N/A&	22.6&	3.9&	2.7&	16&	5.2&	1.5&	3.6&	12&	2.1&	0.7&	3.4&	5.6&	9.3&	1&	2.4&	8.4\\

    \midrule

    Count in CS fields&	[1985-2023]&	18264&	& 	4559&	1078&	1983&	576&	1020&	2129&	643&	1015&	1175&	520&	1640&	968&	2647&	1529&	837&	504\\

    \bottomrule
  \end{tabular}
  }
  }% end makebox
  \label{tab:table3} % Unique label

  \vspace{0.4cm}      % vertical space
\small	Note: Some studies reported results directly as percentages, while others provided raw counts that were converted to percentages for this table. Due to this conversion process, and potential rounding in the original studies, calculated population counts within each study may result in non-integer values (decimals).

\end{table}

\subsection{Data Aggregation}

To synthesize data across all articles, we calculated the raw number of individuals belonging to each MBTI type within each study. This was achieved by multiplying the reported percentage of each MBTI type by the study's total sample size. For instance, if a study reported that 25\% of its 200 participants were ESTJs, we calculated the number of ESTJs as 0.25 × 200 = 50. This calculation was performed for every MBTI type reported in each article, converting percentages into absolute population counts.

\begin{figure} [htbp]
  \centering
    \includegraphics[width=0.9\textwidth]{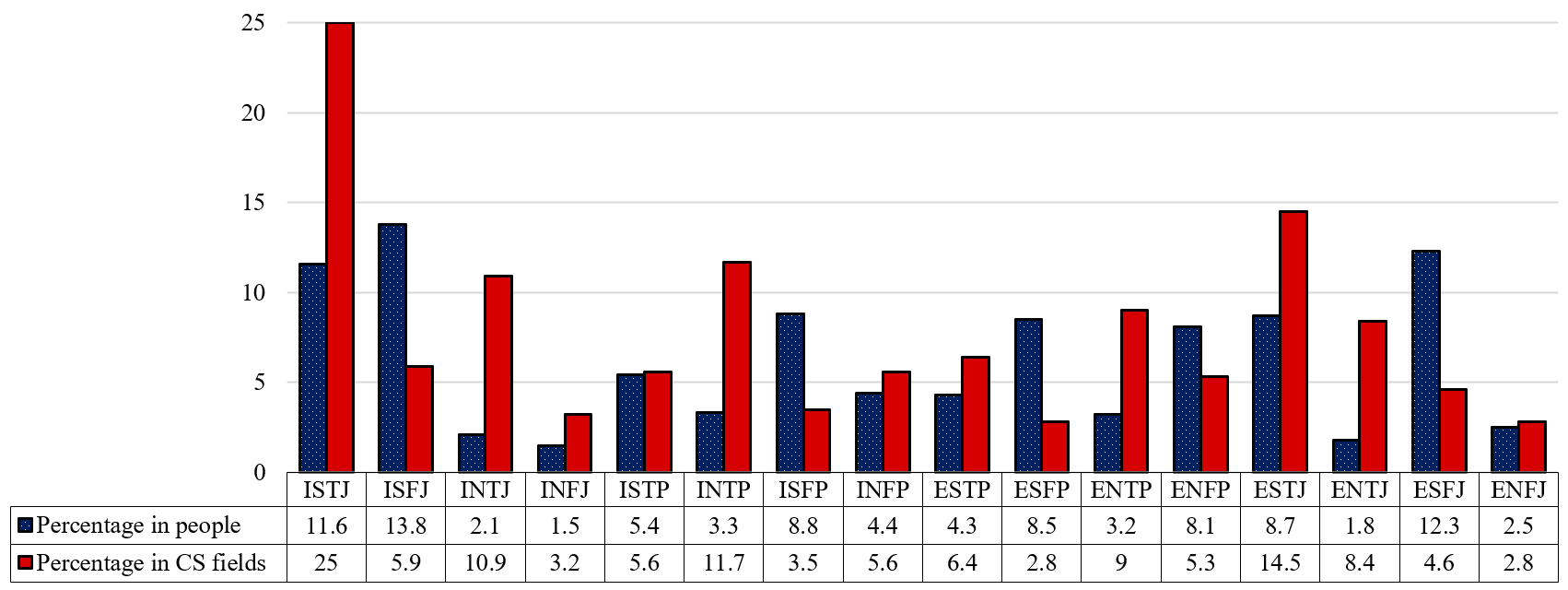} % adjust path and size
  \caption{MBTI Type Distribution in Computer-Related Fields (red) and General Population MBTI Type Distribution (blue) \protect\cite{MBTIDS2017}}    
  \label{fig:fig2} % Unique label

\vspace{0.4cm}      % vertical space

    \small	Important Note: The sum of percentages in this table slightly exceeds 100\% (100.3\%). This discrepancy is due to the original source data \citeA{MBTIDS2017} reporting percentages that, when summed, do not precisely equal 100\%. This is a common occurrence in datasets where percentages are rounded to a limited number of decimal places, leading to minor cumulative rounding errors. These small errors are inherent in the original reported data and do not significantly impact the normalization process.    

\end{figure}

The number of individuals for each MBTI type were then summed across all articles, providing a total count for each personality type (see Table \ref{tab:table3}). To determine the overall percentage representation of each type within computer-related fields, we divided each type's total count by the sum of all types' counts, and then multiplied by 100. This yielded the percentage distribution of MBTI types, as presented in Figure \ref{fig:fig2}.

\subsection{Data Normalization}

To accurately assess the prevalence of MBTI types and cognitive functions within the computer industry, it was essential to account for the baseline distribution of these types within the general population. Simply observing raw frequencies within our computer-related sample could be misleading. For instance, a seemingly high representation of a particular MBTI type in our sample might simply reflect that type's overall commonness in the population, rather than a specific affinity for computer-related careers. Normalization addresses this issue by adjusting the observed frequencies in our sample relative to the expected frequencies based on the general population. This process allows us to identify types and functions that are overrepresented or underrepresented in the computer industry compared to their baseline prevalence. In essence, normalization provides a crucial context for interpreting the raw data, allowing us to determine whether a particular cognitive profile is genuinely more common among computer professionals than would be expected by chance.

Normalization was achieved by dividing the percentage of each MBTI type within computer-related fields by the corresponding percentage in the general population. For example, if 10\% of the general population were ESTJ and 15\% of those in computer-related fields were ESTJ, the normalized value would be 15 ÷ 10 = 1.5. These normalized values were then converted back into percentages by dividing each by the sum of all normalized values and multiplying by 100. This final step ensured that the normalized percentages accurately reflected the relative representation of each MBTI type in computer-related fields. After obtaining the normalized percentages for each MBTI type, we proceeded to analyze the Jungian function stack for each type, as shown in Table \ref{tab:table2}. Each MBTI type is characterized by a specific sequence of cognitive functions, known as the Jungian function stack. For example, ESTJ has the function stack: Te (Extraverted Thinking), Si (Introverted Sensing), Ne (Extraverted Intuition), and Fi (Introverted Feeling).

\section{Prevalence of Functions and Types in Computing}
\label{sec:prevalence}

\begin{figure} [htbp]
  \centering
    \includegraphics[width=0.8\textwidth]{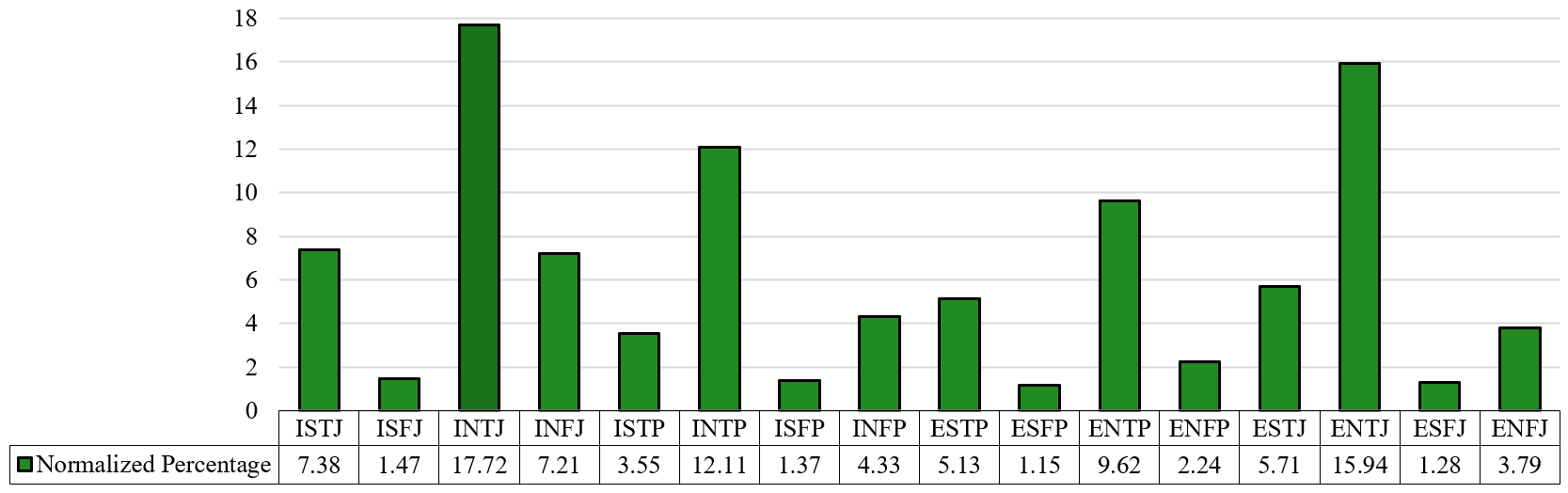} % adjust path and size
  \caption{Normalized MBTI Type Distribution in Computer-Related Fields}
  \label{fig:fig3} % Unique label
\end{figure}

Following the data aggregation and normalization procedures described in Section \ref{sec:methodology}, we present the key findings of our analysis. Figure \ref{fig:fig3} illustrates the normalized distribution of MBTI types within computer-related fields. These percentages represent the relative prevalence of each type after accounting for their distribution in the general population (as shown in Figure \ref{fig:fig2}). A normalized percentage greater than its corresponding value in Figure \ref{fig:fig2} indicates overrepresentation in the computer industry, while a lower percentage indicates underrepresentation.

As Figure \ref{fig:fig3} indicates, certain MBTI types, such as INTJ and ENTJ, exhibit significantly higher normalized percentages, suggesting a strong predisposition for these types to pursue and succeed in computer-related careers. To delve deeper into the underlying cognitive mechanisms driving these preferences, we next examined the distribution of Jungian function pairs.

We focused on the first two Jungian functions of each type, which form pairs that are similar among sister types (types sharing the same cognitive functions but differing in their orientation, introverted vs. extraverted). For instance, ESTJ and ISTJ share the same first two functions, Te and Si, but differ in their extraversion and introversion. By summing the percentages of these sister types, we obtained the total representation for each of the eight possible first two Jungian function combinations within the computer-related fields. These combinations are: Te-Si, Fe-Si, Ne-Ti, Ne-Fi, Te-Ni, Fe-Ni, Se-Ti, and Se-Fi. This aggregation provided a comprehensive overview of how these function pairs are distributed among individuals in the field. We sorted these pairs by their percentage to understand the impact of each combination on pursuing a career in computer-related fields. The results of this analysis are presented in Table \ref{tab:table4}, which illustrates the significance of each function pair.

\begin{table} [htbp]
 \caption{Prevalence of Jungian Function Pairs in Computer-Related Fields}
  \centering
  {\fontsize{8}{10}\selectfont
  \begin{tabular}{
  p{2.1cm}
  p{0.7cm}
  p{0.7cm}
  p{0.7cm}
  p{0.7cm}
  p{0.7cm}
  p{0.7cm}
  p{0.7cm}
  p{0.7cm}
  p{0.7cm}
  }

    \toprule
    \cmidrule(r){1-2}

    First two functions &	Ni Te &	Ti Ne &	Si Te &	Ni Fe &	Ti Se &	Fi Ne &	Si Fe &	Fi Se &	Sum\\

    \midrule

    Percentage&	33.67&	21.73&	13.09&	11.00&	8.68&	6.57&	2.74&	2.52&	100.00\\

    \bottomrule

  \end{tabular}
  }
  \label{tab:table4} % Unique label
\end{table}

Having determined the prevalence of each function pair, our next step was to analyze the individual cognitive functions to understand their influence on pursuing a career in computer-related fields. Each cognitive function, such as Te (Extraverted Thinking), appears in different combinations. For instance, Te appears in both Te-Si and Te-Ni pairs. To quantify the impact of each cognitive function, we aggregated the percentages of all pairs containing that function. For example, for Te, we summed the percentages of the Te-Si and Te-Ni pairs. This process was repeated for each of the eight cognitive functions: Te, Si, Ne, Fi, Fe, Ni, Se, and Ti. Once we had the aggregated values for each cognitive function, we converted these values into percentages. This conversion step transformed the raw counts into percentages, providing a clearer comparison of the relative prevalence of each cognitive function within the field. By sorting these percentages, we could identify which cognitive functions were most prevalent among individuals pursuing careers in computer-related fields. This analysis revealed the dominant cognitive functions and provided insights into their influence on career choices within the field. The results of this detailed analysis are presented in Table \ref{tab:table5}. These findings enhance our understanding of the distribution of cognitive functions among professionals in computer-related fields.

\begin{table} [htbp]
 \caption{Prevalence of Individual Jungian Cognitive Functions in Computer-Related Fields}
  \centering
  {\fontsize{8}{10}\selectfont
  \begin{tabular}{
  p{2.1cm}
  p{0.5cm}
  p{0.5cm}
  p{0.5cm}
  p{0.5cm}
  p{0.5cm}
  p{0.5cm}
  p{0.5cm}
  p{0.5cm}
  p{0.7cm}
  }

    \toprule
    \cmidrule(r){1-2}

    Jungian Functions&	Te&	Ni&	Ti&	Ne&	Si&	Fe&	Se&	Fi&	Sum\\

    \midrule

    Percentage& 23.38&	22.33&	15.20&	14.15&	7.92&	6.87&	5.60&	4.55&	100.00\\

    \bottomrule

  \end{tabular}
  }
  \label{tab:table5} % Unique label
\end{table}

\section{Discussion}
\label{sec:discussion}

Based on the analysis conducted in the methodology section (Section \ref{sec:methodology}) of this study, we have identified and ranked the Jungian functions according to their relative importance and impact on pursuing a career in computer-related fields. This section elucidates the specific role each function plays, contextualized within the demands and nature of computer-related professions. To provide a clear and structured understanding of how Carl Jung's cognitive functions and the MBTI relate to success in computer-related careers, we present the following analysis in a series of tables. This format allows for easy comparison and highlights the key strengths and potential challenges associated with different cognitive preferences. It is important to note why the following tables emphasize the top-ranked individual Jungian functions, dual combinations, and MBTI types. Our primary goal is to identify the cognitive profiles most strongly correlated with success in the computer industry.

Table \ref{tab:table6} presents the four most influential individual Jungian functions, ranked according to their prevalence within computer-related professions. Each function is described and its key strengths are highlighted in the context of specific roles within the industry.

\begin{table} [htbp]
 \caption{Individual Jungian Functions and Their Impact on Computer-Related Careers \protect\cite{jung1971personality}}    
  \centering
  {\fontsize{8}{10}\selectfont
  \begin{tabular}{
  p{1cm}
  p{5.2cm}
  p{5.3cm}
  p{3cm}
  }

    \toprule
    \cmidrule(r){1-2}

    Jungian Function&	Description&	Key Strengths in Computer-Related Fields&	Example Roles\\

    \midrule

    Te&
    Drive to structure, organize, and systematize the external environment through logical and objective methods. Imposes order on complex systems, ensuring efficiency, scalability, and replicability.&
    Project management (defining goals, setting timelines, coordinating teams), software development (creating well-structured code), ensuring alignment of stakeholders with project objectives, creating logical frameworks for initiatives.&
    Project Manager, Software Developer, Systems Analyst\\

    Ni&
    Synthesizes information into a coherent vision, enabling strategic foresight, long-term planning, and innovation. Predicts trends, foresees challenges, and devises comprehensive strategies.&
    Software architecture, systems engineering, technology consulting, AI development (foreseeing model evolution and interaction with real-world data), envisioning future trajectory of projects and technologies.&
    Software Architect, Systems Engineer, Technology Consultant, AI Developer\\

    Ti&
    Driven by internal logic to understand underlying principles. Excels at rigorous analysis, dissecting complex problems, and identifying efficient, elegant solutions. Focuses on internal consistency and logically sound solutions.&
    Algorithm development, debugging, theoretical research, software development, systems engineering, abstracting and conceptualizing complex ideas for research and development.&
    Algorithm Developer, Debugger, Researcher, Software Developer, Systems Engineer\\

    Ne&
    Generates multiple ideas, explores possibilities, and connects seemingly unrelated concepts. Fosters creativity and innovation through divergent thinking.&
    Software design, UX engineering, game development, brainstorming, generating diverse ideas, exploring multiple possibilities, adapting to new and exciting directions.&
    Software Designer, UX Engineer, Game Developer, Innovator\\

    \bottomrule

  \end{tabular}
  }

  \label{tab:table6} % Unique label
\end{table}

While individual functions provide a foundational understanding, the interplay between these functions, forming dual combinations, offers a more nuanced perspective. Certain pairings of functions create synergistic effects, enhancing an individual's aptitude for specific tasks and roles. Table \ref{tab:table7} presents the most impactful dual combinations of Jungian functions. These pairings represent how two primary cognitive functions work together, shaping an individual's approach to problem-solving, innovation, and overall work style within the computer industry. Again, we focus on the top-ranked combinations for clarity and relevance to career success in this field.

\begin{table} [htbp]
 \caption{Dual Combinations of Jungian Functions and Their Impact}
  \centering
  {\fontsize{8}{10}\selectfont
  \begin{tabular}{
  p{1.5cm}
  p{5cm}
  p{5.3cm}
  p{3cm}
  }

    \toprule
    \cmidrule(r){1-2}

    Dual Combination&	Description&	Key Strengths&	Example Roles\\

    \midrule

    Ni-Te&
    Leverages Ni's vision and Te's organization for strategic implementation. Combines long-term vision with actionable, structured outcomes.&
    Strategic planning, efficient execution, driving innovation while maintaining efficiency and results.&
    Software Architecture, Systems Engineering, Technology Management, Leadership Roles\\

    Ti-Ne&
    Combines Ti's logical consistency and deep analysis with Ne's creativity and exploration. Excels at dissecting complex problems and developing novel, innovative solutions.&
    Algorithm development, software engineering, research and development, creating robust and efficient solutions, exploring unconventional approaches, rewarding analytical precision and creative problem-solving.&
    Algorithm Development, Software Engineering, Research and Development\\

    Si-Te&
    Exemplifies reliability, consistency, and methodical execution. Focuses on recalling past experiences and applying established knowledge, paired with logical organization and efficiency.&
    Quality assurance, database management, IT operations, ensuring work is carried out consistently with adherence to standards, maintaining integrity and stability of systems.&
    Quality Assurance, Database Management, IT Operations\\

    Ni-Fe&
    Combines long-term strategic thinking with a deep understanding of interpersonal dynamics. Foresees future trends and develops strategies, ensuring they resonate with stakeholders and align with user needs.&
    Technology consulting, UX design, ethical AI development, integrating technical vision with user experience and ethical considerations.&
    Technology Consulting, UX Design, Ethical AI Development\\

    \bottomrule

  \end{tabular}
  }
  \label{tab:table7} % Unique label
\end{table}

Building upon the understanding of individual functions and their combinations, we can now examine how these elements manifest within the broader framework of the MBTI. Table \ref{tab:table8} presents the top 8 MBTI personality types, ranked in order of their observed prevalence for careers in the computer industry, based on our analysis. It describes the dominant and auxiliary functions of each type, highlighting their key strengths and potential challenges within this field.

\begin{table} [htbp]
 \caption{MBTI Types and Their Suitability for Computer-Related Careers}
  \centering
  {\fontsize{8}{10}\selectfont
  \begin{tabular}{
  p{0.6cm}
  p{2cm}
  p{7.6cm}
  p{4.5cm}
  }

    \toprule
    \cmidrule(r){1-2}

    MBTI Type&	Dominant \& Auxiliary Functions&	Strengths in Computer-Related Fields&	Challenges\\

    \midrule

    INTJ&
    Ni, Te&
    Visionary outlook, strategic planning, integrating complex information into a coherent plan, excelling in roles requiring foresight, innovation, and logical structuring.&
    May struggle with detailed, hands-on tasks if overly focused on the big picture.\\

    ENTJ&
    Te, Ni&
    Natural commanders, thriving in decisive action and strategic vision, organizing and leading projects with a clear structure, setting goals, developing strategies, and mobilizing teams.&
    May be perceived as too direct or blunt, potentially overlooking the importance of interpersonal harmony.\\

    INTP&
    Ti, Ne&
    Deep analytical thought, creative problem-solving, excelling in theoretical domains, dissecting complex problems and generating innovative solutions, exploring underlying principles.&
    May struggle with follow-through due to a tendency to jump from idea to idea.\\

    ENTP&
    Ne, Ti&
    Creativity, innovation, brainstorming, generating a wide range of possibilities, honing ideas into logical solutions, adapting to new situations.&
    May require support from more structured colleagues to ensure follow-through and execution.\\

    ISTJ&
    Si, Te&
    Reliability, attention to detail, methodical approach, excelling in roles requiring precision, consistency, and adherence to protocols, maintaining and applying learned knowledge.&
    May struggle with adapting to rapid changes or unconventional approaches.\\

    INFJ&
    Ni, Fe&
    Visionary thinkers with a deep sense of purpose, long-term planning, considering the human impact of technology, aligning solutions with user needs and values, integrating technical innovation with human values.&
    May struggle with the practical, detail-oriented aspects of implementation.\\

    ESTJ&
    Te, Si&
    Detail-oriented, efficiency driven, management of procedures, and resources&
    May not be as innovative, needs the support of intuitive individuals\\

    ESTP&
    Se, Ti&
    Detail-oriented, adaptable, problem-solving skills, ability to learn things&
    May struggle with theoretical and abstract subjects\\

    \bottomrule

  \end{tabular}
  }
  \label{tab:table8} % Unique label
\end{table}

The MBTI typology provides a comprehensive framework for understanding how different combinations of Jungian functions impact success in computer-related fields. INTJs and ENTJs lead the way with their strategic vision and logical structuring, while INTPs and ENTPs contribute deep analytical thinking and innovative problem-solving. ISTJs and INFJs bring reliability and ethical considerations to the forefront, while ESTJs and ESTPs excel in managing and troubleshooting systems. Each MBTI type offers unique strengths that, when understood and applied effectively, can enhance both individual career development and the overall success of teams within the technology industry. Understanding these typological influences allows for more informed career choices, better team composition, and a more nuanced approach to personal and professional growth in the ever-evolving world of technology. These tables provide a comprehensive, structured framework for understanding the relationship between cognitive preferences, as defined by Jungian functions and the MBTI, and success in the computer industry.

\section{Challenges and Future Perspectives}
\label{sec:challenges}

This study provides insights into the connection between Jungian cognitive functions, MBTI types, and computer industry careers. To ensure a comprehensive understanding and guide future research, several limitations inherent in the scope and methodology warrant consideration. Our meta-analysis incorporates data from diverse studies, spanning various countries, time periods, and potentially different MBTI assessment methods (self-reporting vs. certified practitioners, varying instrument versions). This heterogeneity, along with potential sample biases towards specific subfields (e.g., software engineering), job levels, or demographic groups, and sometimes insufficient sample sizes, may limit the generalizability of our findings. Furthermore, the rapidly evolving nature of the computer industry, with its constant emergence of new roles and technologies, means our results may not fully capture the cognitive demands of future roles or the impact of technological advancements. The correlational nature of our study also prevents us from establishing causal relationships between cognitive profiles and career choices or success; other factors like interests, skills, education, and opportunities undoubtedly play significant roles. Finally, despite including data from multiple countries, significant geographical gaps remain, necessitating further research in underrepresented regions.

To address the limitations of this study and advance understanding in this field, future research should prioritize several key areas. Collecting data from larger, more diverse, and representative samples of computer industry professionals is crucial, encompassing a wider range of roles (beyond software engineering, including IT support, cybersecurity, and data science), experience levels, and geographic locations. Longitudinal studies tracking individuals' career trajectories over time are needed to explore potential causal relationships between cognitive preferences and career success, observing how these functions influence choices, progression, and job satisfaction. Integrating qualitative methods, such as interviews, with quantitative analysis would provide richer context and insights into the lived experiences of individuals. Furthermore, research should focus on the unique cognitive demands of specific roles within the industry (e.g., data scientists, cybersecurity analysts, UX designers) and examine how emerging technologies (AI, machine learning, automation) are reshaping these demands and how individuals with different cognitive preferences, including considerations of their developmental level, adapt. Finally, more extensive cross-cultural comparisons are essential to reveal how cultural factors interact with cognitive preferences in shaping career paths.

\section{Conclusion}
\label{sec:conclusion}

This study investigated the relationship between Jungian cognitive functions, MBTI personality types, and career prevalence within the computer industry. Through a meta-analysis of existing literature and a normalization process accounting for general population distributions, we identified significant overrepresentation of certain cognitive profiles in computer-related fields. Specifically, our findings highlight the prominence of individuals with dominant or auxiliary functions of Extraverted Thinking (Te), Introverted Intuition (Ni), Introverted Thinking (Ti), and Extraverted Intuition (Ne). The dual function combinations of Ni-Te, Ti-Ne, and Si-Te, along with the MBTI types INTJ, ENTJ, INTP, and ENTP, were found to be particularly prevalent. These results underscore the importance of cognitive preferences in shaping career choices and success within the computer industry. The strong presence of Te and Ni suggests that individuals with a drive for logical organization, strategic planning, and long-term vision are particularly drawn to and thrive in this field. The prevalence of Ti and Ne indicates the value of analytical thinking, problem-solving skills, and the ability to generate innovative solutions. While correlation does not equal causation, the significant overrepresentation of these cognitive profiles, even after normalization, strongly suggests a meaningful link between personality and career fit within the computer industry.

The practical implications of these findings are considerable. For individuals considering a career in the computer industry, understanding their own cognitive strengths and preferences can inform career path selection and development. For organizations, this research provides valuable insights for talent acquisition, team formation, and management strategies. By recognizing the cognitive diversity within teams and aligning individuals' strengths with appropriate roles, companies can foster greater collaboration, innovation, and overall productivity. Furthermore, educational institutions can utilize this information to better guide students interested in computer-related fields, providing tailored support and resources based on their cognitive profiles. While limitations exist, as discussed previously, this study contributes to a deeper understanding of the interplay between personality and career success in the dynamic and ever-evolving computer industry. The findings provide a foundation for future research to further explore the causal relationships, refine the methodologies, and expand the scope to encompass a broader range of roles, technologies, and cultural contexts. Ultimately, recognizing and leveraging the diverse cognitive strengths of individuals is crucial for both personal fulfillment and organizational success in the rapidly changing landscape of the tech sector.

% Step 4: Use \printbibliography to generate the list
% \printbibliography
%Bibliography
% \bibliographystyle{unsrt}  
\bibliography{references} 
\bibliographystyle{apacite}

\end{document}